# Multipurpose S-shaped solvable profiles of the refractive index: application to modeling of antireflection layers and quasicrystals

## J.-C. KRAPEZ

*ONERA, The French Aerospace Lab, DOTA, F-13661 Salon de Provence, France*
*krapez@onera.fr*

A class of four-parameter solvable profiles of the electromagnetic admittance has recently been discovered by applying the newly developed Property & Field Darboux Transformation method (PROFIDT). These profiles are highly flexible. In addition, the related electromagnetic-field solutions are exact, in closed form and involve only elementary functions. In this paper, we focus on those that are S-shaped and we provide all of the tools needed for easy implementation. These analytical bricks can be used for high-level modeling of lightwave propagation in photonic devices presenting a piecewise-sigmoidal refractive-index profile such as, for example, antireflection layers, rugate filters, chirped filters and photonic crystals. For small amplitudes of the index modulation, these elementary profiles are very close to a cosine profile. They can therefore be considered as valuable surrogates for computing the scattering properties of components like Bragg filters and reflectors as well. In this paper we present an application for antireflection layers and another for 1D quasicrystals (QC). The proposed S-shaped profiles can be easily manipulated for exploring the optical properties of smooth QC, a class of photonic devices that adds to the classical binary-level QC.

## 1. INTRODUCTION

The classical method for modeling light propagation in a continuously heterogeneous dielectric film (see Fig. 1 for a description of the canonical case) consists in slicing it into thin homogeneous sublayers and then applying the well-known analytical transfer matrix approach [1-3]. However, this method is approximate; it is all the more precise when the discretization step, in terms of optical thickness, is small with respect to the considered free-space wavelength $\lambda$. As a rule of thumb, to achieve acceptable numerical results, the optical thickness steps $n_i \Delta z_i$ should typically be smaller than $\lambda/60$ [4]. Replacing these elementary steps of *constant* index by *graded* profiles that are *analytically solvable* (i.e., for which a closed-form analytical solution to Maxwell's equations is known) would provide the double benefit of reducing the number of discretization sublayers and generating a synthetic model that may be continuous, or even of a higher differentiability class, with respect to position.

Analytical solutions, sometimes with the associated transfer matrices, have been proposed for the following profile functions: linear, exponential, power law, sinusoidal, hyperbolic cosine, hyperbolic tangent, and Epstein profile (these functions describe either the refractive index or the permittivity) [5-17]. In all these cases, the computation of *special functions* is required, namely Bessel, Mathieu, Airy, Hermite, Heun or hypergeometric functions. However, solutions based on *elementary functions* should be preferred for reasons of computation time, especially if repeated evaluations are necessary, such as for inverse scattering modeling. Known solvable profiles that lead to solutions involving only elementary functions are few: a four-parameter refractive index profile of this type was presented in [18] and a series of algebraic or 2nd-order polynomial profiles were described in [19, 20]. The latter profiles are concave or convex and they can be stitched together to produce a continuous composite profile. However, since they are defined with only three parameters, there are serious limitations when trying to obtain a composite profile with the first derivative continuous at all nodes (not to mention the second derivative). We have to mention the possibility of generating sequences of 1D *periodic* index profiles, together with the related electromagnetic (EM)-field functions, all based on trigonometric functions only [21-25].

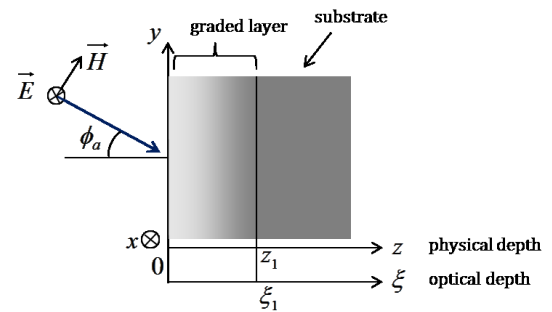

**Fig. 1.** Geometry considered for the case of a graded layer laid over a homogeneous substrate (canonical case considered in § 2-3). The graded layer extends from 0 to the physical depth $z_1$, which corresponds to the optical depth $\xi_1$. The case of a plane wave with TE polarization and incidence angle $\phi_a$ is represented here. In § 4-5, multiple graded layers will be stacked together.

In [26] we reported a new method based on joint PROperty and FIeld Darboux Transformations (the dubbed "PROFIDT method") for building sequences of solvable profiles of the EM *tilted optical admittance* in the *optical-thickness* space. It simultaneously yields the closed-form expressions of the related EM fields $E$ and $H$ (for both polarization modes TE and TM). The method is devoted to media whose permittivity $\varepsilon$ and permeability $\mu$ are real-valued, positive and show continuous (or at least piecewise-continuous) variations along one direction, say $z$ (this $z$-dependency will be omitted in the following expressions, except when really needed). The *tilted optical admittance* is defined by $\eta \equiv \sqrt{\varepsilon/\mu} \cos^{m_p} \phi$, where $\phi$ is the local incidence angle; $m_p = +1$ for the TE polarization mode and $m_p = -1$ for the TM mode. However, in the present paper, we will restrict our discussion to nonmagnetic materials ($\mu = \mu_0$, where $\mu_0$ is the permeability of free space). Hence, the tilted admittance $\eta$ reduces to $Y_0$ times the (refractive) *pseudoindex* $n^*$, with $n^* = n \cos^{m_p} \phi$, where $n$ is the $z$-dependent refractive index $n \equiv \sqrt{\varepsilon/\varepsilon_0} \equiv \sqrt{\varepsilon_r}$ and $Y_0 \equiv \sqrt{\varepsilon_0/\mu_0}$ is the free-space admittance [1]. Since they are related with a constant factor, all developments made in [26] for the tilted admittance $\eta$ can be readily translated for the pseudoindex $n^*$, which will be the case in the sequel. On the other side, the *effective* (or tilted) optical thickness (or optical depth) $\xi$, as measured along the $[0, z]$ geometrical-depth interval (see Fig. 1), is defined by:

$$\xi = \int_0^z n(u) \cos \phi(u) du. \quad (1)$$

A preliminary Liouville transformation changes Maxwell's equations (in the physical-depth space) into Schrödinger equations (in the optical-depth space) for the transformed electric and magnetic scalar fields. The square root of the admittance (here the pseudoindex), resp. its reciprocal, obeys the same Schrödinger equation but with a zero eigenvalue. Successive Darboux transformations can then be applied to obtain chains of solvable pseudoindex profiles together with the related EM fields [26]. Among the infinite sequences of pseudoindex profile solutions provided by the PROFIDT method, one class of 4-parameter profiles is particularly interesting. It is described through a function $s(\xi)$ that represents either the square root of the pseudoindex $n^*$ or its reciprocal. The function $s(\xi)$ is a linear combination (LC) of two independent solutions $B(\xi)$ and $D(\xi)$ of a 2nd order differential equation satisfied by $n^{*+1/2}$, resp. $n^{*-1/2}$ ($A_B$ and $A_D$ are two arbitrary constants):

$$s(\xi) \equiv \left[n^*(\xi)\right]^{\frac{1}{2}m_f} = A_B B(\xi) + A_D D(\xi)$$
$$\begin{cases} B(\xi) = \operatorname{sech}(\hat{\xi}) \\ D(\xi) = \sinh(\hat{\xi}) + \hat{\xi} \operatorname{sech}(\hat{\xi}) \end{cases} \quad ; \quad m_f = \pm 1. \quad (2)$$

In Eq. (2), the argument $\hat{\xi}$ results from a linear transformation of the optical thickness $\xi$ according to the following expression (for ease, a centered formula is preferred here):

$$\hat{\xi}(\xi) = \frac{\xi_1}{\xi_c}\left(\frac{\xi}{\xi_1} - \frac{1}{2}\right) + \gamma, \quad (3)$$

where $\xi_1$ is the optical thickness of the considered graded layer (i.e., over the interval $[0, z_1]$), $\xi_c$ and $\gamma$ are two additional free-parameters that may take values in $]0, +\infty[$, resp. $]-\infty, +\infty[$. In Eq. (2), $\operatorname{sech} = 1/\cosh$ is the hyperbolic secant function. The class of profiles generated by the LC in Eq. (2) was thus dubbed of "$\operatorname{sech}(\hat{\xi})$-type". In Eq. (2), two sub-classes of profiles of the pseudoindex $n^*(\xi)$ are actually represented: the first one, as obtained by setting the exponent of $n^*(\xi)$ to $+1/2$, stems from the Helmholtz equation expressed for the electric field $E$ ("$\langle E \rangle$-form" profiles; $m_f = +1$); the second one, as obtained by setting the exponent to $-1/2$, stems from the Helmholtz equation expressed for the magnetic field $H$ ("$\langle H \rangle$-form" profiles; $m_f = -1$). All of these profiles are defined with *four* parameters: $A_B$, $A_D$, $\gamma$ and $\xi_c/\xi_1$ (the current optical thickness $\xi$ and the "characteristic" optical thickness $\xi_c$ will be systematically non-dimensionalized by the optical thickness of the graded layer $\xi_1$, see Eq. (3)). Notice that the latter two parameters $\gamma$ and $\xi_c/\xi_1$ act *non-linearly* in the definition of $s(\xi)$ (see Eq. (2) and Eq. (3)).

In fact, we came to the same function as the one reported in Eq. (2) in a previous paper devoted to heat diffusion in graded media [27]. Therein, the linear combination $A_B B(\xi) + A_D D(\xi)$ was used to describe a class of solvable profiles of the property: $s(\xi) = b^{\pm 1/2}(\xi)$, where $b$ is the (graded) thermal *effusivity* and $\xi$ represents the *square root of the heat diffusion time* along the path $[0, z]$. The same analytical tools can be used to model temperature fields and EM fields in graded media, which is quite noticeable. In this respect, let us mention Ref. [28], where interesting connections have been highlighted between heat diffusion on one side and EM field attenuation in a specific class of metals on the other side.

Numerous numerical trials have led to the conjecture in [27] that these 4-parameter $\operatorname{sech}(\hat{\xi})$-type profiles present an unbounded *flexibility*. By this, we mean that they can satisfy any set of four specifications regarding the two end-values and the two end-slopes of the leading property of the graded layer. A few illustrative examples were provided in [26] with a set of refractive-index profiles satisfying different combinations of end-slopes. Actually *two solutions* have been systematically obtained: an $\langle E \rangle$-form profile and an $\langle H \rangle$-form profile.

In this paper, we will focus on the sub-class of $\operatorname{sech}(\hat{\xi})$ profiles with *horizontal end-slopes*, i.e., rising or descending S-shaped profiles. We dubbed them ZESST profiles (Zero-End-Slope $\text{Sech}(\hat{\xi})$-Type profiles). Three potential applications were already outlined in [26]. First, they provide an easy model for index-matching layers; a rapid comparison was made with the classic quintic profile [4, 29]. Next, a locally-periodic profile was built by stitching together alternately rising and falling ZESST profiles, for the purpose of modeling an apodized rugate filter (rugate profiles are efficient solutions for producing notch filters deprived of ripples and side-lobes [30-37]). Finally, a model for chirped mirrors (as used for ultrafast lasers [38, 39]) was produced by slowly varying the width and amplitude of the assembled ZESST profiles.

In this paper, we aim to build upon the previous study and analyze the specific features of the ZESST profiles in greater depth. In Section 2, we will provide practical tools for their design. In Section 3, we will consider the design of smooth index-matching layers and antireflection coatings and will describe the results of a comparative analysis with other profiles from the literature. Section 4 will be devoted to an improved matching layer design: it is obtained by joining three $\text{sech}(\hat{\xi})$-type profiles and it is continuous up to the second derivative. Section 5 provides another example of ZESST profile applications, namely the modeling of *smooth quasi-periodic multilayers*. It specifically addresses the transmission properties of Fibonacci quasicrystals, in particular the photonic bandgap structures. Section 6 is a discussion about other potential applications and a conclusion.

## 2. PROFILES OF $\text{sech}(\hat{\xi})$ TYPE WITH S-SHAPE

### A. Profile construction

Let us consider a graded layer extending from $z = 0$ to $z = z_1$ in the physical space, which correspond to $\xi = 0$ and $\xi = \xi_1$ in the optical-depth space (see Fig. 1). In the first part of the paper, this graded layer is simply bounded by two homogeneous and semi-infinite media: on the left, an incident medium with refractive index $n_a$ (typically air) and on the right, a substrate with refractive index $n_s$. In the second part of the paper, multiple graded layers will be stacked together.

A plane wave is impinging from the left side with an incidence angle $\phi_a$ in the incident medium. The incidence angle at optical depth $\xi$ is $\phi(\xi)$. Basic relations between $n^*$ and $n$ (see Annex A) together with the relation between $n^*$ and the function $s(\xi)$ in Eq. (2) provide the necessary tools for translating any boundary specification (regarding level or slope) on either $n(\xi)$, $n^*(\xi)$ or $s(\xi)$ into equivalent boundary specifications on the two other parameters (obviously, for the ZESST profiles considered here, the zero-end-slope specifications are common to all three parameters). Then, having at hand two boundary specifications on $s(\xi)$, namely $s(\xi = 0) = s_0$ and $s(\xi = \xi_1) = s_1$, and two others on its derivative, i.e. $s'(\xi = 0) = s'_0$ and $s'(\xi = \xi_1) = s'_1$, the four parameters $\xi_c/\xi_1$, $\gamma$, $A_B$ and $A_D$ must be evaluated by solving a system of four equations that are linear in $A_B$ and $A_D$, but *non-linear* in $\xi_c/\xi_1$ and $\gamma$. Although a standard non-linear root-finding solver can provide the solution in a relatively short time, it would be desirable to be able to do without. For this reason, an alternative method is now proposed for the ZESST profiles, which consists in evaluating the two non-linear parameters $\xi_c/\xi_1$ and $\gamma$ with the empirical relations in Eqs. (4) and (5). They have been determined after having performed some numerical tests leading to the following observations: $n_0^*$ and $n_1^*$ intervene only through their ratio $n_1^*/n_0^*$; moreover, switching from a value of this ratio to its reciprocal induces nothing other than a sign change to $\gamma$; the same happens when switching from an $\langle E \rangle$-form profile to an $\langle H \rangle$-form profile. A closer analysis when the index ratio $n_1^*/n_0^*$ approaches one value, revealed that $\xi_1/\xi_c = x^{1/2}(a_0 + O(x))$ and $|\gamma| = \text{Arcsinh}(1 + O(x))$ where $x = |\ln(n_1^*/n_0^*)|^{2/3}$, $a_0$ is the first term in the left column in Table 1, and $O(x)$ means a term of order $x$. Then, the terms $O(x)$ have been fitted with polynomials of the variable $x$ to obtain:

$$\xi_1/\xi_c = x^{1/2} \sum_{j=0}^{J} a_j x^j, \quad \textbf{(4)}$$

$$\gamma = m_f \cdot \text{sgn}(n_0^* - n_1^*) \cdot \text{Arcsinh}\left[1 + x \sum_{j=0}^{J} b_j x^j\right], \quad \textbf{(5)}$$

The coefficients $a_j$ and $b_j$ are reported in Table 1 for $J=3$.

**Table 1. Coefficients of the polynomial fittings in Eqs. (4) and (5)**

| j | $a_j$ | $b_j$ |
|---|---|---|
| 0 | 1.6188704 | 1.0912949·10⁻¹ |
| 1 | 1.0061871·10⁻¹ | 1.0813551·10⁻¹ |
| 2 | 2.6169621·10⁻² | 2.3368282·10⁻² |
| 3 | 6.3935152·10⁻³ | 7.5773489·10⁻³ |

When performing the fitting, a quite large domain was considered for $n_1^*/n_0^*$, far beyond the values encountered in optics; actually, the proposed empirical relations are intended to be applicable to a broader class of problems, including for example acoustics, microwaves and transmission lines. The relative error on $\xi_c/\xi_1$, resp. the absolute error on $\gamma$ (after being scaled by $\xi_1/\xi_c$) is less than 3·10⁻⁵, resp. 8·10⁻⁶ when the index ratio $n_1^*/n_0^*$ is within the range [0.1, 10]. In this range, the RMS difference between the "exact" ZESST profile $n^{*\,m_f/2}(\xi)$ and that inferred from the fitted values of $\xi_c/\xi_1$ and $\gamma$ (after scaling by $\left|n_1^{*\,m_f/2} - n_0^{*\,m_f/2}\right|$) is less than 2·10⁻⁵.

Once the two parameters $\xi_c/\xi_1$ and $\gamma$ are determined from Eqs. (4) and (5), the remaining two parameters $A_B$, $A_D$ are simply inferred from the two equations expressing the boundary conditions on $s(\xi)$, namely regarding $n_0^{*\,\frac{1}{2}m_f}$ and $n_1^{*\,\frac{1}{2}m_f}$ (by use of Eq. (2)). This finally solves the determination of the profile $s(\xi)$ and thereby the pseudo-index profile $n^*(\xi)$, which is straightforwardly inferred from $n^*(\xi) = s^2(\xi)$ or $n^*(\xi) = s^{-2}(\xi)$, depending on whether the $\langle E \rangle$-form case or the $\langle H \rangle$-form case is being considered.

In Fig. 2, we plotted a series of ZESST-profiles for a relatively large range of end-to-end index-ratio values, namely from 0.25 to 4. Both the $\langle E \rangle$-form and $\langle H \rangle$-form profiles are reported. In the present case of ZESST profiles, these two forms are very close to each other; a much greater difference is generally observed in the situation of slanted end-slopes – see [26].

One should stress the following point. The PROFIDT method provides successive sets constituted of a pair of *pseudoindex profiles* $n^*(\xi)$ (i.e., $\langle E \rangle$-form and $\langle H \rangle$-form), together with the corresponding analytical solutions for the EM fields $E$ and $H$, for both the TE and TM modes (the related transfer matrices will be presented thereafter).

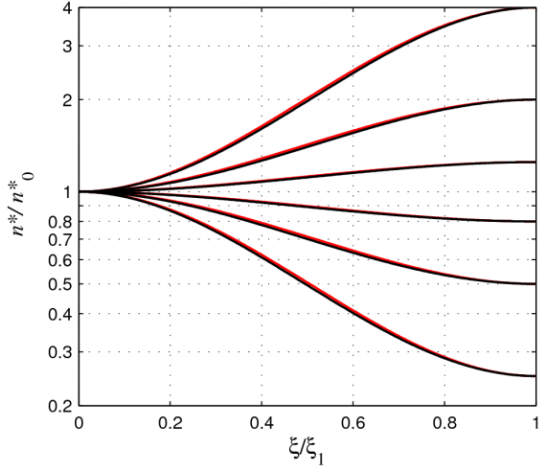

**Fig. 2.** Solvable profiles of $\text{sech}(\hat{\xi})$-type with zero-end-slope (ZESST profiles) for the *pseudoindex* $n^*(\xi)$ ($n^*(\xi)$ is divided by the value taken at the left end of the layer, i.e., $n_0^*$). They are expressed against the normalized optical thickness $\xi/\xi_1$, where $\xi_1$ is the total optical thickness of the graded layer. Six values are considered for the right-to-left refractive-index ratio $n_1^*/n_0^*$: 0.25, 0.5, 0.8, 1.25, 2 and 4. In black: profiles of $\langle E \rangle$-form, in red: profiles of $\langle H \rangle$-form (they nearly overlap).

Then, given an incidence index $n_a$ and an incidence angle $\phi_a$ (hence a Snell-Descartes invariant $I_a$ - see Annex A), any such pseudoindex profile $n^*(\xi)$ can be translated into an *index profile* $n(\xi)$ with the help of Eq. (A1) for TE polarization, resp. Eq. (A2) for TM polarization. For this reason, one can argue that the *solvable profiles* expressed in terms of refractive index $n(\xi)$ are $I_a$-*dependent*, whereas the original ones, i.e., those regarding the pseudoindex $n^*(\xi)$, are not. Consequently, the EM field solutions are valid for a refractive-index profile $n(\xi)$ and the sole $I_a$ value that was used in Eq. (A1) or Eq. (A2) to infer it. If the EM fields had to be computed for any other incidence-angle value (more precisely, any other $I_a$ value), one should refer to other methods, like the classical (homogeneous) transfer matrix, and apply it to discretized homogeneous slices.

If desired, a last step can be applied for changing from the optical-thickness space $\xi$ back into the geometrical-thickness space $z$, which corresponds to an inverse Liouville transformation. The underlying operations were fully described in [26]. We will merely summarize the procedure that is applicable for normal incidence or, in case of oblique incidence, for TE polarization only. In these cases, the profiles of $n^*$ or $n$ can be analytically expressed vs. the geometrical depth $z$, albeit in *implicit form only*. This is based on the following relationship between $\xi$ and $z$ (whereby $\xi$ spans the interval $[0, \xi_1]$ and $z$ spans the interval $[0, z_1]$):

$$z(\xi) = \xi_c \cdot (f_{E,H}(\xi) - f_{E,H}(0)), \quad (6)$$

$$f_E(\xi) = \begin{cases} \dfrac{-1}{2A_D} \dfrac{B(\xi)}{n^{*1/2}(\xi)} & \text{if } A_D \neq 0 \\ \text{or } \dfrac{1}{2A_B} \dfrac{D(\xi)}{n^{*1/2}(\xi)} & \text{if } A_B \neq 0, \end{cases} \quad (7)$$

$$f_H(\xi) = A_B^2 \tanh(\hat{\xi}) + A_D^2 \left( \dfrac{1}{4} \sinh(2\hat{\xi}) - \dfrac{\hat{\xi}}{2} + \hat{\xi}^2 \tanh(\hat{\xi}) \right) \quad (8)$$
$$+ 2A_B A_D \hat{\xi} \tanh(\hat{\xi}).$$

In the case of TM polarization, the inverse Liouville transformation involves a quadrature that cannot be reduced to a closed-form analytical expression. The transformation $\xi \leftrightarrow z$ should then be performed numerically (see [26]). For this reason, in the sequel, examples of the inverse Liouville transformation will concern normal incidence or, in case of oblique incidence, TE polarization only.

With this restriction in mind, applying the inverse Liouville transformation described in Eqs. (6)-(8) on the ZESST profiles in Fig. 2 yields the profiles plotted in Fig. 3.

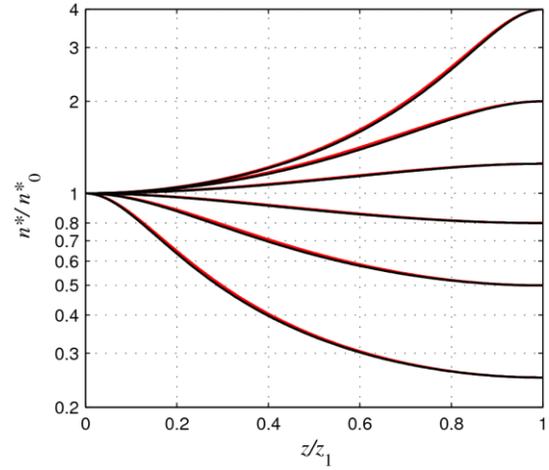

**Fig. 3.** ZESST profiles of Fig. 2 plotted against the normalized physical depth $z$ ($z_1$ is the total thickness of the graded layer). Normal incidence or oblique incidence with TE polarization was assumed for performing the inverse Liouville transformation from $\xi$ to $z$.

**B. Surrogate models for modulated profiles (periodic and almost periodic)**

Normalizing the ZESST profiles according to $(n^* - n_0^*)/(n_1^* - n_0^*)$ provides another perspective on their shape; see Fig. 4 and Fig. 5. For reference, we also plotted a normalized cosine profile.

For falling profiles (i.e., $n_1^*/n_0^* < 1$), the profile curvature in Fig. 4 and Fig. 5 is higher at the left end of the layer, whereas for rising profiles, it is the opposite. This means that the curvature is more pronounced towards the boundary presenting the highest index value. This dissymmetry is more acute in the $z$-space (Fig. 5) than in the $\xi$-space (Fig. 4). Furthermore, when the amplitude of the modulation diminishes (i.e., when $n_1^*/n_0^*$ approaches 1), the $\text{sech}(\hat{\xi})$-type profiles, both of $\langle E \rangle$-form and $\langle H \rangle$-form, increasingly resemble a cosine profile. One can even notice that the $\langle H \rangle$-form profiles are slightly closer to it (for a given value of $n_1^*/n_0^*$). The RMS difference between a

ZESST profile and the cosine profile gives a measure of their similarity. Interestingly, the RMS difference is the same for a given value of $n_1^*/n_0^*$ and for its reciprocal. For example, if we refer to the two pairs of curves in Fig. 5 that correspond to $n_1^*/n_0^*$ =0.8 and 1.25, namely the closest ones to the cosine profile, the RMS difference is 0.06 for both $\langle E \rangle$-type (black) profiles, and only 0.04 for both $\langle H \rangle$-type (red) profiles.

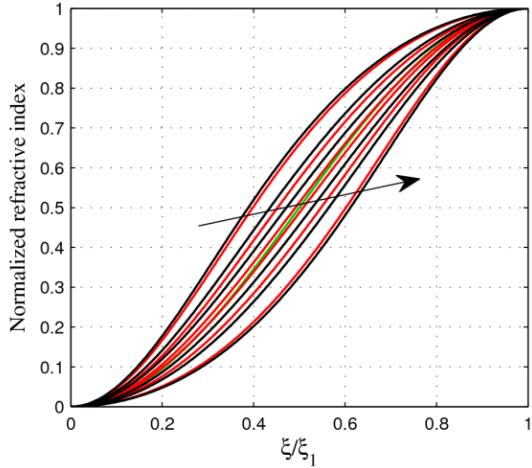

**Fig. 4.** Same as in Fig. 2 for the normalized pseudoindex $(n^*-n_0^*)/(n_1^*-n_0^*)$. The arrow indicates increasing values of the right-to-left index ratio $n_1^*/n_0^*$ : 0.25, 0.5, 0.8, 1.25, 2, 4. The normalized cosine profile has been added in green.

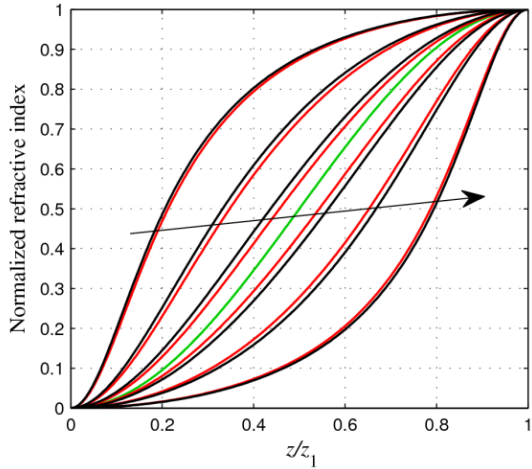

**Fig. 5.** Same as Fig. 4 against the normalized physical depth $z/z_1$. Normal incidence or oblique incidence with TE polarization.

More insight into this similarity is provided in Fig. 6, where the RMS difference is plotted against $|\ln(n_1^*/n_0^*)|$. This variable was chosen to place greater emphasis on low-contrast profiles. The data for rising profiles (i.e., $n_1^* > n_0^*$) and for falling profiles ($n_1^* < n_0^*$) obviously overlap. Except for $\langle H \rangle$-form profiles and a relatively low contrast, i.e., $|n_1^*/n_0^* - 1| < 0.05$, the proximity to the cosine profile is better in the $\xi$-space than in the $z$-space (this was already noticed when comparing Fig. 4 and Fig. 5). Moreover, we can notice that for a vanishing contrast (i.e., $n_1^*/n_0^* \to 1$) the RMS difference has a non-vanishing limit of about 0.007, which, anyway, is quite low.

Let us fix the RMS acceptance-threshold at $10^{-2}$: profiles showing a lower RMS difference value in the $z$-space will be considered as satisfactory *surrogate models* for sine/cosine profiles. The $\langle E \rangle$-form profiles meet this criterion provided that $n_1^*/n_0^* \in$ [0.988; 1.012]. For $\langle H \rangle$-form profiles, the allowable interval is much broader: $n_1^*/n_0^* \in$ [0.932; 1.073]. After doubling the acceptance threshold, the allowable interval for $n_1^*/n_0^*$ broadens to [0.95; 1.05] for $\langle E \rangle$-form profiles, resp. to [0.88; 1.13] for $\langle H \rangle$-form profiles, which is now quite large. In the end, for applications dealing with an index contrast not higher than 5-12% (as for Bragg filters and reflectors) and requiring only 1-2% accuracy on the actual profile shape, the ZESST profiles can advantageously replace the classic sine-cosine profiles (the benefit is that the former involve elementary functions for computing the EM fields and scattering properties, whereas the latter involve Mathieu functions [5, 6]).

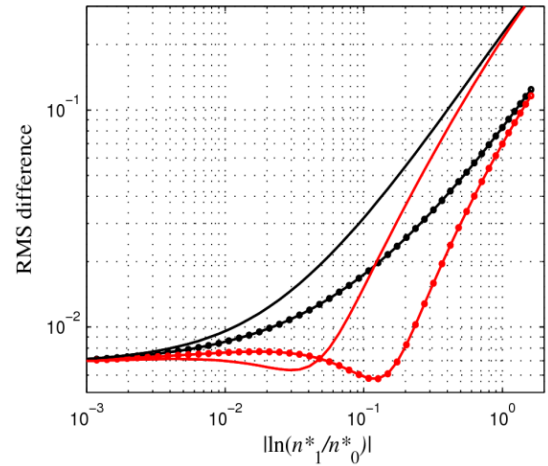

**Fig. 6.** RMS difference between the normalized ZESST profiles and the cosine profile (see Fig. 4 and Fig. 5) against the logarithm of the pseudoindex contrast $n_1^*/n_0^*$ (in absolute value). Dotted curves: RMS difference in the $\xi$-space (see Fig. 4), plain curves: same in the $z$-space (see Fig. 5). In black: profiles of $\langle E \rangle$-form, in red: profiles of $\langle H \rangle$-form.

### C. Transfer matrix

The analytical transfer matrix is used for expressing the linear relationship between the vector of tangential components of the EM fields at two positions in a graded material, namely $[E_1, \tilde{H}_1]^t$ at $\xi = \xi_1$ and $[E_0, \tilde{H}_0]^t$ at $\xi = 0$ :

$$\begin{bmatrix} E_0 \\ \tilde{H}_0 \end{bmatrix} = \mathbf{M} \begin{bmatrix} E_1 \\ \tilde{H}_1 \end{bmatrix}, \quad \mathbf{M} = \begin{bmatrix} A & B \\ C & D \end{bmatrix}, \quad \textbf{(9)}$$

where $E$ represents $E_x$ or $E_y$ in the case of the TE mode, resp. the TM mode. Symmetrically, $\tilde{H}$ represents $H_y/Y_0$ or $H_x/Y_0$.

Transfer matrices related to different profiles can then be multiplied in sequence for modeling light propagation through the corresponding stacking, in the same way as for homogeneous layers [1, 2, 26]. The scattering properties (reflectance, transmittance) of the resulting synthetic optical structure can be inferred from the four entries of the global matrix, together with the refractive indices of the incidence medium and the substrate through well-known relationships (see e.g. [1, 2]).

We will designate by $\mathbf{M}_{\langle E \rangle}$ and $\mathbf{M}_{\langle H \rangle}$ the transfer matrix associated with an $\langle E \rangle$-form, resp. an $\langle H \rangle$-form, profile. The generic expressions of the four entries of the matrix $\mathbf{M}_{\langle E \rangle}$ related to a solvable profile in the $\xi$-space were developed in [26] and they are recalled in Annex B. The interesting point is that $\mathbf{M}_{\langle H \rangle}$ is obtained simply by a 180° circular permutation of the corresponding $\mathbf{M}_{\langle E \rangle}$ matrix [26].

Assuming an EM plane wave (either TE or TM-polarized), with $\lambda$, the wavelength in free-space, and $k_0 = 2\pi/\lambda$, the wavenumber, the tangential component of the electric field, i.e., $E_x$ for the TE mode, or $E_y$ for the TM mode inside a $\text{sech}(\hat{\xi})$ profile of $\langle E \rangle$-form is expressed as $n^{*-1/2}(\xi)$ times a linear combination involving the following function:

$$K(\xi, k_0) = \left(i\sqrt{k_0^2 \xi_c^2 - 1} - \tanh(\hat{\xi})\right) \exp\left(i\sqrt{k_0^2 \xi_c^2 - 1}\, \frac{\xi}{\xi_c}\right), \quad \textbf{(10)}$$

and its complex conjugate $P(\xi, k_0) = \overline{K}(\xi, k_0)$ [26]. Symmetrically, the magnetic-field related to the $\langle H \rangle$-form profiles (i.e., $H_y$ for the TE mode, resp. $H_x$ for the TM mode) is expressed as $n^{*+1/2}(\xi)$ times a linear combination involving the same functions $K(\xi, k_0)$ and $P(\xi, k_0)$. The matrices related to the $\text{sech}(\hat{\xi})$-type profiles are obtained by substituting the expressions of the linearly independent functions $K(\xi, k_0)$ and $P(\xi, k_0)$ presented in Eq. (10). Skipping the intermediate analytical details, we provide the results in synthetic form in Annex B. All that is needed to calculate the four entries of $\mathbf{M}_{\langle E \rangle}$ or $\mathbf{M}_{\langle H \rangle}$ is fully contained in Equations (B1) and (B3). One additional useful result is that the transfer matrices of two symmetric ZESST profiles evolving from $n_0^*$ to $n_1^*$, resp. from $n_1^*$ to $n_0^*$, differ only by an exchange of the entries A and D.

## 3. APPLICATION OF ZESST PROFILES TO MATCHING LAYERS AND AR COATINGS

A first application that springs to mind is the design of gradient-index matching layers or antireflection coatings. Reflections originating at the interface of two dissimilar media with index $n_0$ and $n_1$ may be significantly reduced over a broad spectral range by the use of an intermediate layer with a smooth transition between the two index values [4, 29, 40-42]. A continuous first derivative at both ends of the matching layer is obtained with the cubic function $n = n_0 + (n_1 - n_0)(3t^2 - 2t^3)$ [4] and with the cosine function $n = n_0 + (n_1 - n_0)[1 - \cos(\pi t)]/2$. Therein, the argument $t$ represents either the normalized geometrical thickness $z/z_1$ or the normalized optical-thickness $\xi/\xi_1$. With the quintic function $n = n_0 + (n_1 - n_0)(10t^3 - 15t^4 - 6t^5)$, the second derivative is continuous too and Southwell showed that the antireflection properties of the quintic profile are improved compared to those of the cubic one; it is near optimum [4, 29]. Nanorod layers of TiO$_2$ and SiO$_2$ were grown by oblique angle deposition to produce gradual-index multilayers approximating the quintic profile [43-44]. Another type of matching layer is obtained with the hyperbolic tangent profile $n^2 = n_0^2 + (n_1^2 - n_0^2)[\tanh(\kappa_1(t - 1/2)) + 1]/2$, where $\kappa_1$ is used for adjusting the steepness of the transition (it belongs to the family of Epstein layers, see [15]). Broadband omnidirectional antireflection (AR) coatings were claimed in [45] by adding to the former tanh profile $-V(t)/k_{AR}^2$, where $V(t)$ is one among the well-known reflectionless potentials (RP) described by Kay and Moses [46] and $k_{AR} = 2\pi/\lambda_{AR}$ is the AR-design wavenumber (they will be referred to as "tanh+RP" profiles). Notice that since both tanh and RP profiles were originally designed for unbounded media (infinite support), a balance has to be found between the opposing effects of steepness and support truncation.

The ZESST profiles may offer an interesting alternative, since: 1- they allow a continuous first derivative to be obtained at the layer ends, and 2- the computation of the associated EM solution and the scattering properties is analytical, exact and easy (refer to Annex B).

A comparison of the ZESST profiles of $\langle E \rangle$-form and of $\langle H \rangle$-form with the classical profiles mentioned previously is shown in Fig. 7 and Fig. 8, where the scaled *index* $n/n_0$ is plotted against the scaled optical depth $\xi/\xi_1$, resp. the scaled physical depth $z/z_1$. For this illustration, we considered an index-step ratio of 1.5. Two profiles were drawn for the cubic, cosine and quintic functions, depending on the choice for the argument $t$, i.e., either $t = \xi/\xi_1$ (plain curves) or $t = z/z_1$ (dashed curves). A profile in $t = \xi/\xi_1$ is symmetric about the center at $\xi/\xi_1 = 1/2$ and $n/n_0 = (1 + n_1/n_0)/2$, see Fig. 7. The same is observed in Fig. 8 after interchanging $\xi/\xi_1$ and $z/z_1$.

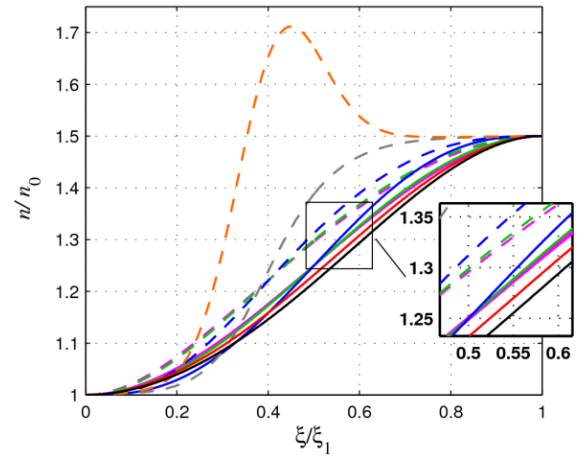

**Fig. 7.** Index profiles considered as gradient-index matching layers between two media with index contrast $n_1/n_0 = 1.5$. The scaled index is plotted against the normalized optical depth $\xi/\xi_1$. Black: ZESST profile of $\langle E \rangle$-form, red: profile of $\langle H \rangle$-form, magenta: cubic profiles, green: cosine profiles, blue: quintic profiles, grey: tanh profile, orange: "tanh+RP" profile. Continuous (resp. dashed) lines: analytical profiles expressed in $t = \xi/\xi_1$ (resp. in $t = z/z_1$).

For the tanh profile and the "tanh+RP" profile [45], we used $t = z/z_1$ and $\kappa_1 = 7$, which provides a not-too-steep ramp and, at the same time, insignificant discontinuity at the boundaries. With regard to the reflectionless potential, we used the simplest one, namely the classic hyperbolic secant potential (also known as modified Pöschl-Teller potential), $V(t) = -2\kappa_2^2 z_1^{-2} \operatorname{sech}^2 \kappa_2(t-1/2)$, which is here centered at $t = 1/2$ (we selected an AR design wavelength of $\lambda_{AR} = n_0 z_1/2$ and a parameter $\kappa_2 = 10$ – the maximum contrast induced by the RP is then about $0.5 \, n_0$).

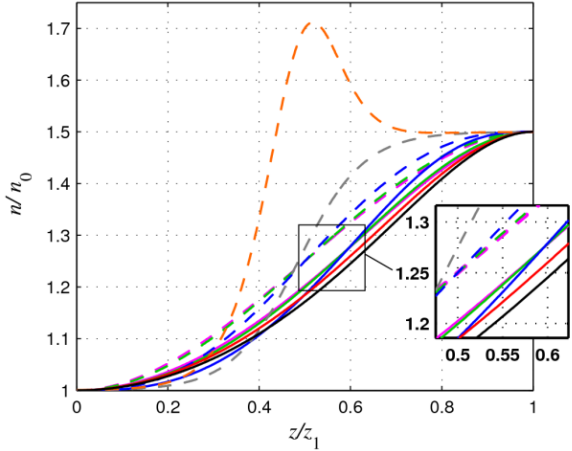

**Fig. 8.** Same as Fig. 7 against the normalized physical depth $z/z_1$.

Assuming a plane wave coming from the left in Fig. 1, the reflectance spectra for the ten matching layers represented in Fig. 7-8 are reported in Fig. 9-10 in logarithmic scale and in Fig. 11 in linear scale. For all, except the two solvable ZESST profiles, we had to implement the classical homogeneous-layer analytical transfer matrix method. As such, in order to lower the error induced by this approximation, we imposed the conservative constraint $\Delta \xi_i < \lambda/120$, which implies a discretization into about 600 sublayers for safely exploring the considered spectral band, i.e., down to a wavelength of $\lambda = \xi_1/5$. Analyzing shorter wavelengths would require discretizing even more densely. Instead of that, the analysis of each ZESST profile required computing only one single matrix, according to the methodology described in Annex B. Although a bit more complicated than the transfer matrix of a homogeneous layer, this single matrix is processed in much less time than is required to obtain the (approximate) one related to the classical profiles (i.e., as obtained after multiplying a series of constant-index matrices resulting from the spatial discretization).

The reflectance spectra can be split into three groups, depending on the continuity/differentiability properties of the profiles. In Fig. 9 are presented those pertaining to $C^1$ and $C^2$ profiles, i.e., those with a continuous first derivative (cubic, cosine and ZESST profiles), resp. those with a continuous second derivative (quintic profiles). In Fig. 10 are plotted the spectra of the discontinuous profiles (tanh profile and "tanh+RP" profile). These profiles are $C^\infty$ over an infinite support but here the support had to be truncated to $[0, z_1]$.

For vanishing values of $\lambda^{-1}$ we retrieve, for all profiles, the well-known Fresnel reflection value for a bare interface (0.04 in the present case). Thereafter, all spectra show a global decrease of reflectance for increasing $\lambda^{-1}$ (except for the "tanh+RP" spectrum with an overshoot at $n_0 z_1/\lambda \approx 0.5$). The reflectance decrease is more or less rapid, depending on the index profile shape.

It appears that the reflectance spectra of the $C^1$ and $C^2$ profiles in Fig. 9 are significantly lower than those of the third group in Fig. 10. In particular, due to the support truncation of the latter group, and hence the appearance of a refractive-index discontinuity, their spectra show ripples and cease to decrease for $n_0 z_1/\lambda$ higher than about 3. In Fig. 10 and Fig. 11 is also plotted the reflectance spectrum of the Kay-Moses RP profile alone, i.e., without the tanh ramp. The perfect reflectionless condition is met for $n_0 z_1/\lambda = 2$, which corresponds to the design AR wavelength. However, on both sides of this depression, the reflectance is quite high (with respect to the other spectra). Next, two important observations can be made about adding a tanh ramp to the Kay-Moses RP: first, the reflectionless feature is erased; secondly, the resulting "tanh+RP" spectrum is *higher everywhere* than the underlying tanh spectrum. In addition, it is higher than any other AR profile considered in this paper (either the classic ones or the ZESST profiles).

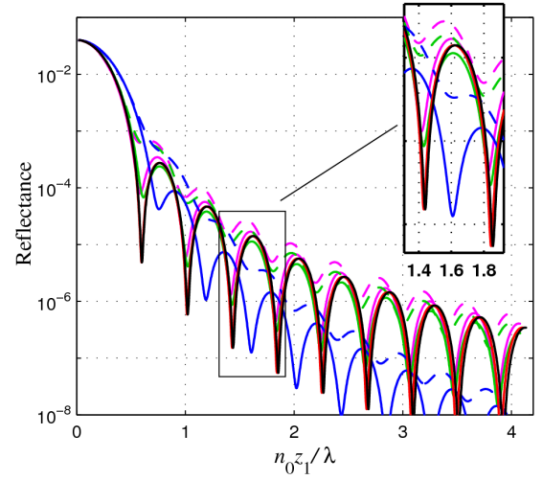

**Fig. 9.** Reflectance spectra of the $C^1$ and $C^2$ profiles reported in Figs. 7-8 (same colors and line types) against the scaled reciprocal wavelength $n_0 z_1/\lambda$. The spectra of the ZESST profiles of $\langle E \rangle$-form (black) and $\langle H \rangle$-form (red) are very close and show deep depressions alternating with those of the quintic profile in $t = \xi/\xi_1$ (plain curve in blue).

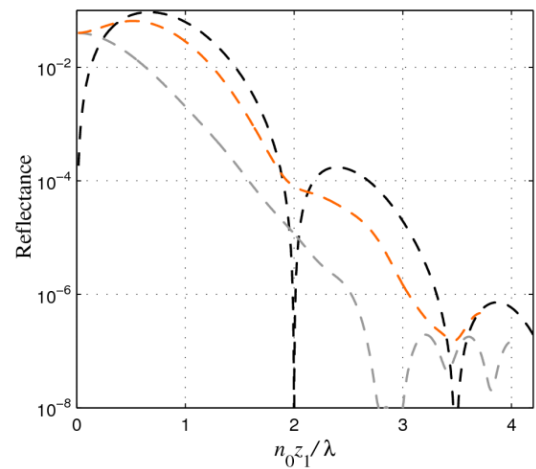

**Fig. 10.** Reflectance spectra of the tanh (grey) and "tanh+RP" (orange) profiles in Figs. 7-8; a black dashed line is added, which corresponds to the spectrum of the sech² reflectionless potential (RP) alone (i.e., without the tanh ramp); the deep depression in the central area corresponds to the perfectly reflectionless condition at the AR design wavelength $\lambda_{AR}/n_0 z_1 = 0.5$.

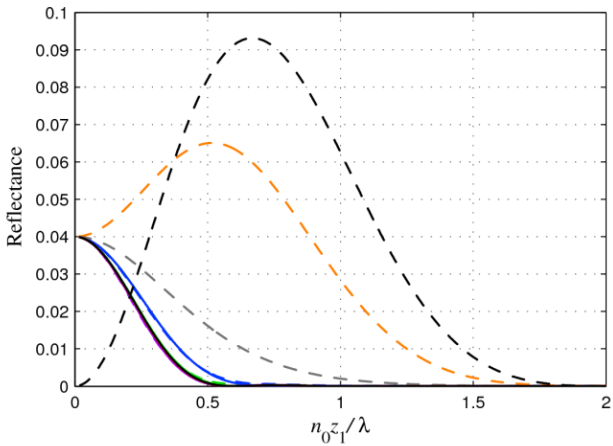

**Fig. 11.** Reflectance spectra of Figs. 9 and 10 reproduced in a linear scale (same colors and line types). Disregarding the dashed black line, which simply recalls the result of the plain sech² "reflectionless" potential (RP), the other spectra correspond, in decreasing order of anti-reflection performances, to the "tanh+RP" profile (dashed orange), the "tanh" profile (dashed grey), both quintic profiles (dashed and plain blue) and a group formed of (both) cubic, (both) cosine and (both) $\mathrm{sech}(\hat{\xi})$-type profiles. For a better view of the performances at reflectance levels lower than about 10⁻³, one should refer to Figs. 9 and 10.

For the three classic profiles: cubic, cosine and quintic, a better reduction of reflectance is achieved by choosing the normalized optical depth $\xi/\xi_1$ (continuous curves) instead of the physical depth $z/z_1$ (dashed curves) for the functional parameter $t$. In addition, the periodic minima are more pronounced. This preference for $t = \xi/\xi_1$ was already noticed by Southwell for the quintic profiles [29].

The spectrum of the ZESST profiles is always (slightly) lower than that of the cubic profile expressed in $\xi/\xi_1$. It is actually closely comparable to that of the cosine profile expressed in $\xi/\xi_1$: the reflectance maxima of the ZESST profiles are slightly higher, however, the minima are much deeper.

In Fig. 9, the spectra of the $\langle E \rangle$-form and $\langle H \rangle$-form ZESST profiles are very close, since the profiles themselves are very close (see Fig. 7 or Fig. 8). One interesting point is that when the wavelength is scaled by the total optical thickness (instead of $n_0 z_1$ in Fig. 9), these spectra *perfectly overlap* (this phenomenon is also discussed in Section 4). We are in the presence of two refractive-index profiles that, although distinct, lead to the same reflectance, whatever the considered wavelength. Other pairs of similarly *spectrally indistinguishable profiles* were described in [26].

Interestingly, the reflectance minima observed with the quintic profiles roughly correspond to $\xi_1/\lambda \approx j/2$, $j \geq 2$, whereas for all other profiles (except for the tanh and the tanh+RP profile) they are localized at $\xi_1/\lambda \approx (2j-1)/4$, i.e., nearly a quarter-wave apart.

Disregarding the local minima, the global decrease of reflectance with decreasing wavelength soon stabilizes at a rate of 10⁻⁴/decade for the first group and 10⁻⁶/decade for the second group. These results are consistent with the general trend reported in [40, 42, 47, 48]: in the presence of a $C^{j-1}$ transition profile (i.e., when the $(j-1)^{th}$ derivative is continuous but not the $j^{th}$), the reflectance spectrum is expected to evolve like $\lambda^{2j}$ for vanishing $\lambda$ (this relationship should, however, not be extrapolated for $j \to \infty$, as pinpointed in [48]). As such, the steeper decrease of the quintic profile reflectance spectrum in Fig. 9 would give an advantage to this particular profile for designing AR coatings. Nevertheless, this statement should be tempered since for $n_0 z_1/\lambda$ <0.7 the ZESST profiles show better results than the quintic profile; thereafter the performances alternate, and for $n_0 z_1/\lambda$ >0.9 the reflectance is less than 10⁻⁴ for all three profiles anyway. In practice, the choice must then be made by jointly considering the spectral range of interest and the allowable thickness for the AR layer.

Another point to consider is the influence of the incidence angle. Assuming again an index step $n_1/n_0$ of 1.5, the variation with the incidence angle of the ZESST profile spectra of the mean reflectance is described in Fig. 12 in logarithmic scale and in Fig. 13 in linear scale. The results for $\phi_a$ =0° were obtained by implementing a single "high-level" transfer matrix, as described in Annex B. For the other angle values, as discussed earlier in § 2.A, it was necessary to implement the classical transfer matrix method with a fine discretization of the ZESST profiles drawn in Fig. 8.

As seen in Figs. 12-13, the $\langle E \rangle$-form profile provides slightly better results than the $\langle H \rangle$-form profile. In the former case, a reflectance lower than 1% is reached from normal to 60° incidence provided that $n_0 z_1/\lambda$ >0.678, which means for wavelengths shorter than 1.47 $n_0 z_1$. The wavelength should be shorter than 0.7 $n_0 z_1$ (i.e., $n_0 z_1/\lambda$ >1.44) to obtain a reflectance lower than 0.1% over the same incidence range. In order to obtain less than 1% reflectance over the incidence range [0°-70°], the wavelength should be shorter than 0.64 $n_0 z_1$ (i.e., $n_0 z_1/\lambda$ >1.56). This gives some indications for the design of ZESST-type omnidirectional AR coatings.

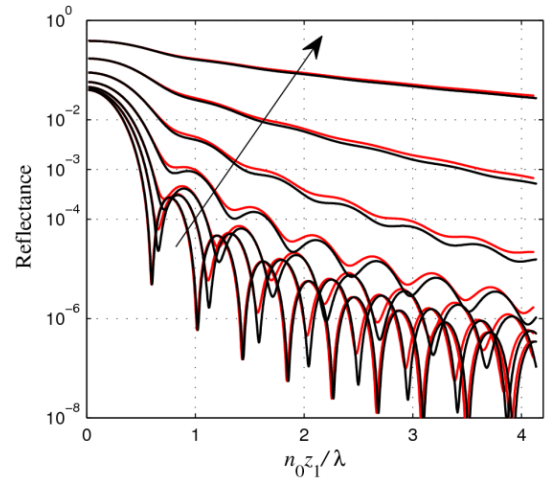

**Fig. 12.** Mean reflectance spectra $(R_{TE} + R_{TM})/2$ of the ZESST profiles in Fig. 8 depending on the incidence angle $\phi_a$ ($\langle E \rangle$-form (black) and $\langle H \rangle$-form (red)). From bottom to top: $\phi_a$ =0°, 30°, 40°, 50°, 60°, 70°, 80°. Index step $n_1/n_0$ =1.5.

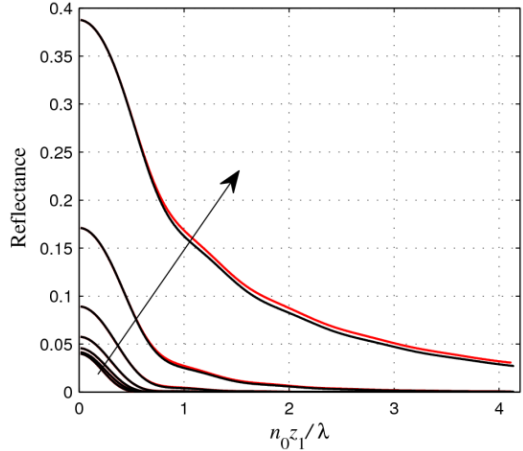

**Fig. 13.** Same as Fig. 12 in a linear scale.

## 4. COMPOSITE $\text{sech}(\hat{\xi})$-TYPE PROFILES FOR $C^2$ MATCHING LAYERS

The slower global reduction of the ZESST profile reflectance with a decreasing wavelength, as compared to the quintic profiles (see Fig. 9), was explained by the fact that the former are continuous up to the *first derivative* only, whereas the latter are continuous up to the *second derivative*. Stitching together *several* $\text{sech}(\hat{\xi})$-type profiles offers a chance to build a solvable composite profile of better performance than a single ZESST profile. The objective is now to build a composite profile that would be continuous up to the second derivative at all nodes. Joining *two* $\text{sech}(\hat{\xi})$-type profiles is not enough since we then have nine function specifications and only 2x4=8 free parameters. Joining *three* $\text{sech}(\hat{\xi})$-type profiles is a feasible solution, since we then have twelve function specifications and as many as 3x4=12 free parameters. Furthermore, we have two degrees of freedom left for assigning the relative positions of the two internal nodes. A specific routine has been developed to identify the twelve unknown parameters by fusing a nonlinear solver devoted to six of them with a direct identification of the remaining six (linear) parameters.

In Fig. 14 we describe the results obtained when distributing the two internal nodes evenly, i.e., at $\xi/\xi_{total}$ =1/3 and 2/3 (to avoid any misinterpretation, the notation $\xi_{total}$ is chosen to describe the optical thickness of the whole matching layer, whether it is a 1-piece or a 3-piece layer). The dashed curves of the two composite profiles (in black: for the $\langle E \rangle$-form and in red for the $\langle H \rangle$-form) can be compared with the former single-piece ZESST profiles and with the quintic profile.

The reflectance spectra are reported in Fig. 15. As opposed to Fig. 9, the wavelength is now scaled by the total optical thickness. The $\langle E \rangle$-form and $\langle H \rangle$-form 3-piece profiles give *strictly the same reflectance spectrum* (just as the $\langle E \rangle$-form and $\langle H \rangle$-form 1-piece ZESST profiles do) Complementary computations (not shown here) revealed that this is not the case when the two internal nodes are set asymmetrically with respect to the middle point.

Fig. 15 confirms that the reflectance spectra of the 3-piece $\text{sech}(\hat{\xi})$-type profiles have the desired reduction rate of $10^{-6}$/decade. In the present case of equidistant nodes, we can observe ripples of high amplitude and width 1 (in $\xi_1/\lambda$ units) alternating with ripples of low amplitude and width 0.5. These low-amplitude ripples extend over quite large spectral bands, where the reflectance is exceptionally low (as opposed to the narrow reflectance minima observed with the other profiles, i.e., the quintic and the ZESST profiles). Nevertheless, antagonist effects can be noticed when comparing with the 1-piece ZESST profiles: better performances are (globally) reached with the 3-piece $\text{sech}(\hat{\xi})$ profiles at short wavelengths (typically for $\lambda < \xi_{total}$), whereas the opposite is observed at long wavelengths ($\lambda > \xi_{total}$).

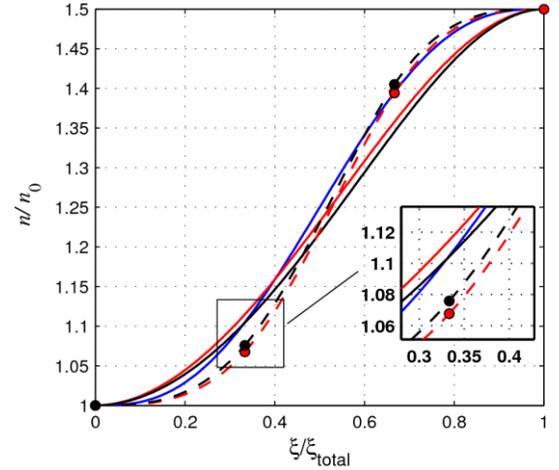

**Fig. 14.** In dashed lines: $C^2$ composite profiles obtained by joining three $\text{sech}(\hat{\xi})$-type profiles; the circles indicate the connection nodes (in black: $\langle E \rangle$-form, in red: $\langle H \rangle$-form profiles). For comparison, we reproduced three curves from Fig. 7: continuous line in black: single ZESST profile of $\langle E \rangle$-form, in red: ZESST profile of $\langle H \rangle$-form, in blue: quintic profile expressed in $t = \xi/\xi_1$.

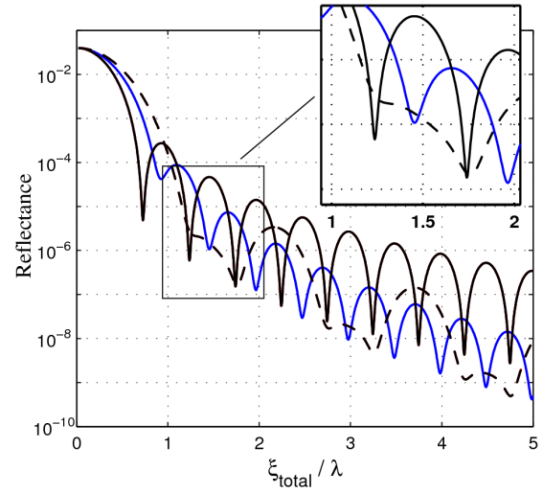

**Fig. 15.** Reflectance spectra of the profiles in Fig. 12 (same colors and line types). The reflectance spectra of the $\langle E \rangle$-form and $\langle H \rangle$-form profiles overlap, whether they are from the 3-piece composite profiles (black dashed line) or from the 1-piece ZESST profiles (black continuous line).

## 5. SMOOTH 1D QUASICRYSTALS

1D QC are structures made of layers arranged using well-designed patterns with long-range order, but lacking translational symmetry [49, 50]. An aperiodic distribution of refractive index variations induces optical interferences, which, when compared to their periodic counterparts (photonic crystals), yield richer and more complex features in the transmission spectrum. Over the past thirty years, a large number of studies have been devoted to the exploration and exploitation of the interference peculiarities offered by Fibonacci quasicrystals and other deterministic aperiodic structures like Thue-Morse, Rudin-Shapiro and period-doubling sequences (see e.g. [51-54] and the reviews [49, 50]). The aperiodic structures are generally obtained by applying specific substitution rules on two building blocks, say $A$ and $B$. In almost all previous works, $A$ and $B$ correspond to *homogeneous* layers defined by their indexes $n_A$, $n_B$ and optical thicknesses $\xi_A$, $\xi_B$. As a result, the index profile was a binary-level profile with a *discontinuity* at each $AB$ or $BA$ interface. The research has been focused on the interplay between aperiodic sequences and optical scattering properties, in particular distinctive resonant states with various degrees of spatial confinement. Typical features are localized optical states (i.e., Anderson-like states) and pseudo-gaps separated by strongly fluctuating wavefunctions with power-law localization scaling, known as critical modes. These critical modes include extended fractal wavefunctions and result in self-similar spatial fluctuations [50, 54]. Another intriguing property of aperiodic multilayers is the appearance of perfect transmission resonances in the optical spectra (i.e., transmittance is exactly equal to unity) [55]. Nevertheless, only a small number of works took into consideration index profiles other than binary-level profiles. Namely, in [5, 56-59], one layer type, say $A$, has been changed to a graded-index layer (with linear or exponential profile); in [7] both layer types have been changed to graded-index layers. However, in all of these cases, the resulting profiles were still discontinuous.

It is well known that the scattering properties of a photonic structure depend on the index variation amplitude and space scales. Therefore, the interplay between the Fourier spectrum of an aperiodic lattice and its energy spectrum has been the subject of intense research [49, 50]. The differentiability order of a profile also has an impact on reflectance [42], an aspect which, to the best of our knowledge, has not yet been considered in the field of quasicrystals. For this reason, we propose to generate *smooth* 1D quasicrystals using ZESST profiles and compare the optical response with that of their *discrete* counterparts, i.e., binary-level quasicrystals. The present illustration will be about deterministic aperiodic structures based on Fibonacci sequences.

Fibonacci sequences are obtained by applying the following iteration rule: $S_n = [S_{n-1} S_{n-2}]$, initialized with $S_1 = [A]$ and $S_2 = [AB]$. At fifth order, for example, we obtain $S_5 = [ABAABABA]$. For the *discrete*-QC, we assign, as usual, the same optical-thickness value to both layers: $\xi_A = \xi_B = \xi_1$. The *smooth* QC is constructed as follows: each $AB$ (resp. $BA$) transition is "smoothed" and replaced by a ZESST profile of optical thickness $\xi_1$ evolving from $n_A$ to $n_B$ (resp. from $n_B$ to $n_A$). Wherever a $AA$ block is present, a layer of index $n_A$ and optical thickness $\xi_1$ is inserted between the two neighboring ZESST profiles. Both discrete-QC and smooth-QC are assumed to be bounded by infinite layers of index $n_A$ (notice that to avoid an index discontinuity at the right boundary if the sequence ended with $B$, a supplementary ZESST profile from $n_B$ to $n_A$ is added – this occurs for any even order in the Fibonacci sequence). The discrete and smooth profiles can be compared in Fig. 16.

To compute the scattering properties of the discrete-QC, two elementary transfer matrices, one for each layer type, $A$ and $B$, have to be prepared and multiplied according to the Fibonacci sequence. For the proposed smooth-QC, one needs to compute two transfer matrices as well: one for the homogeneous layer $A$ and one for the rising ZESST profile (the matrix of the falling ZESST profile is obtained by exchanging A and D entries in the latter matrix).

In Fig. 17, we present the transmittance spectra for both discrete and smooth QC sequences at 12$^{th}$ order ($S_{12}$) with chosen refractive indices of $n_A$=1.6 and $n_B$=2.2 ($\langle E \rangle$-form profiles were used for the ZESST profiles; those of $\langle H \rangle$-form provide very close results). The spectra are plotted over a wavenumber interval extending by 10% on each side of the characteristic wavenumber, for which both layers $A$ and $B$ are quarter-wave.

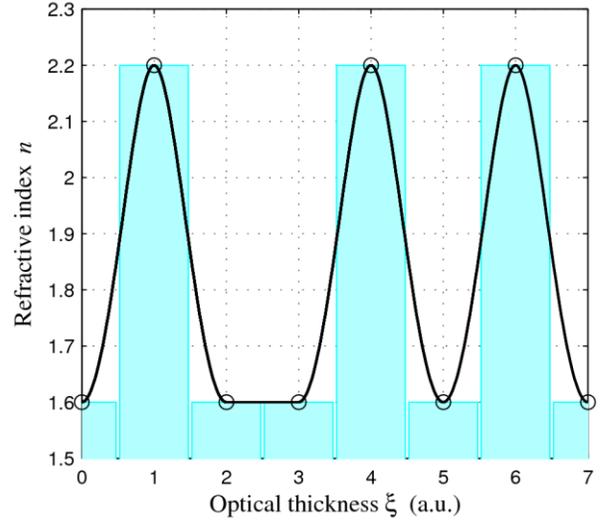

**Fig. 16.** In blue: schematic diagram showing the left part of an aperiodic lattice composed of two layers $A$ and $B$, arranged in a Fibonacci sequence ($n_A$=1.6, $n_B$=2.2, same optical thickness for both layers, which is used as the $\xi$-unit). A *smooth* profile is obtained by substituting a rising (resp. falling) ZESST profile of $\langle E \rangle$-form at each $AB$ (resp. $BA$) interface. The circles indicate the nodes of the resulting smooth quasicrystal.

At first glance, the spectrum of the smooth profile presents the same features as the discrete counterpart, i.e., many pseudo-gaps separated by critical modes showing a multifractal scaling with narrow transmission lines. By increasing the sequence order, we observe, in the same way as with homogeneous layers, a deepening of the pseudo gaps and the appearance of new narrow spectral features. Actually, quasi-localization of the light waves in a Fibonacci dielectric multilayer was demonstrated by the self-similarity of the transmission coefficient [52]. The main difference that can be observed in Fig. 17 is a slight shift of the features towards higher wavenumbers (when using $\langle H \rangle$-form profiles instead -not represented here-, the shift is just slightly more pronounced).

A more striking difference with the discrete-QC spectrum is observed for a *reduced wavenumber* $\Omega = 4\xi_1/\lambda$ higher than 2 (i.e., for a wavelength $\lambda$ smaller than $2\xi_1$), see Fig. 18. In the spectral region corresponding to $\Omega \in [1.7, 2.3]$, the transmittance reaches high levels for both quasicrystals. Thereafter, the transmittance of the discrete QC again enters a perturbation region, which is very similar to the pseudo-gap and multifractal region left from $\Omega=2$. Actually, the discrete-QC spectrum shows a periodicity of 2 units in $\Omega$ and a symmetry about any integer value of $\Omega$, as was highlighted in [51].

On the contrary, the transmittance of the smooth QC remains at high levels for $\Omega>2$, showing only a few narrow dips. The depth of these dips increases with the sequence order. $\Omega=2$ corresponds to $\lambda=2\xi_1$, i.e., a wavelength equal to the spatial period of the photonic crystal obtained by suppressing the $AA$ doublings in the Fibonacci sequences (these supernumerary $A$ layers can be seen as pseudo-random "defects"). Hence, for $\lambda<2\xi_1$, the smooth quasicrystals are essentially transparent, notwithstanding some sparse and thin stopbands. Only for wavelengths greater than $2\xi_1$ do they behave much like the classical two-level stepwise Fibonacci sequences.

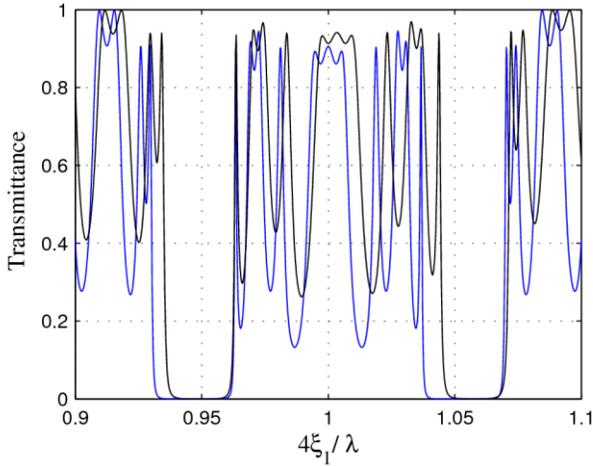

**Fig. 17.** Transmittance spectra of the binary-level Fibonacci sequence $S_{12}$ (i.e., 233 layers) (in blue) and its smooth counterpart based on ZESST profiles of $\langle E \rangle$-form (in black). $n_A$ =1.6 and $n_B$ =2.2. The abscissa $4\xi_1/\lambda$ is the reduced wavenumber, where $\xi_1$ is the common optical-thickness value for all elements.

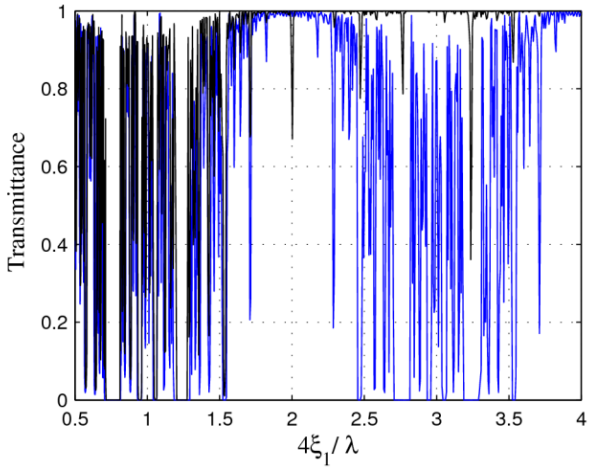

**Fig. 18.** Same as Fig. 17 over a larger spectral domain.

The strong difference in transmittance observed at high wavenumbers between the binary-level profile and the smooth profile should be related to the fact that the Fourier spectrum of the binary-level profile has a higher content at high frequency than that of the smooth profile. In the end, we can notice that the transmittance spectrum of the smooth QC has lost all of the symmetry properties of the discrete QC counterpart.

## 6. DISCUSSION AND CONCLUDING REMARKS

We have described the main features of a pair of S-shaped refractive index profiles, together with their exact EM analytical solutions. These so-called ZESST profiles (Zero-End-Slope $\text{Sech}(\hat{\xi})$-Type profiles) are a particular subclass of a more general class of solvable profiles, the $\text{sech}(\hat{\xi})$-type profiles, which were obtained in [26] by applying the PROFIDT method (Darboux transformation method). For any value of the right-to-left index ratio of the S-shape, two solvable profiles are actually proposed: the so-called $\langle E \rangle$-form and $\langle H \rangle$-form profiles. They are defined in the optical-depth space and are aimed at modeling the EM fields for both the TE and TM modes. The EM-field expressions are exact, in closed-form and involve only elementary (hyperbolic/trigonometric) functions.

Two of the four parameters that define each ZESST profile are non-linear; however, empirical relations have been provided for easy but accurate determination. Practical tools are available to manage an analytical representation back in the physical-depth space (this applies for TE polarization; otherwise, the inverse Liouville transformation should be performed numerically). Analytical formulas have also been given to calculate the corresponding transfer matrices. All ingredients are thus available for the computation of the scattering properties of one or several S-shaped profiles bound together. Let us emphasize that this computation is exact.

In this paper we explored the performances of ZESST profiles when used as matching (or antireflection) layers. They compare favorably with other profile solutions from the literature, in particular with the well-known quintic profiles. One needs to compute one single transfer matrix for the ZESST profile, as opposed to a multiplicity of them when dealing with the other matching-layer profiles, since they require the application of the classical analytical transfer matrix over very fine homogeneous layers. Thus, implementing ZESST profiles eliminates both the burden with the discretization-step criterion and the round-off error problem induced by the fine discretization. The AR performance at high wavenumber can be increased further by changing from a single-piece ZESST profile to a 3-piece $\text{sech}(\hat{\xi})$ profile, since the latter shows greater smoothness (i.e., it is continuous up to the 2nd derivative). Nevertheless, if the reflectance target for the AR coating at normal incidence is not less than $3 \cdot 10^{-4}$, a single ZESST profile provides about the best solution among all of the profiles considered here: for any wavelength shorter than about $1.9\, n_0 z_1$, the reflectance would be less than the aforementioned threshold (these numerical results were obtained for a 50% index step).

The 3-element composite profile that has been considered in this paper is the first example of what we coined a "*solvable $\text{sech}(\hat{\xi})$-type spline*". Since the $\text{sech}(\hat{\xi})$-type profiles are 4-parameter flexible functions, they could well be used in spline interpolation in lieu of the classical 3rd degree polynomials (cubic spline). The great advantage lies in that each $\text{sech}(\hat{\xi})$-type profile element is (exactly) solvable and that any combination thereof is (exactly) solvable too: through a simple transfer-matrix multiplication one has access to the scattering properties of the whole synthetized profile.

Solvable *pseudo-splines* are obtained by relaxing the constraint on the continuity of the second-derivative at the nodes (only the first derivative should be continuous). Joining together alternately rising and falling ZESST profiles with a progressively changing width and/or height yields such pseudo-splines. This gives rise to almost-periodic (solvable) index-profiles that can be used to model a huge number of optical devices, like apodized rugate filters, fiber Bragg filters and mirrors, chirped mirrors and photonic crystals (a brief introductory outline of such applications was provided in [26]). It has been shown in this paper that ZESST profiles, especially those of $\langle H \rangle$-form, are very satisfactory substitution models for sine/cosine profiles of low to moderate amplitude, which is the case with Bragg filters, among others. The mean discrepancy with the cosine function can be less than 1%, which is often well acceptable.

With regard to the application to 1D photonic quasicrystals, the results presented in this paper provide a first glimpse of the opportunities offered by the ZESST profiles for the analysis of lightwave propagation in *smooth* quasicrystals. As a matter of fact, a photonic device with locally periodic and smooth variations of the refractive index can be easily modeled with ZESST profiles, as we have shown with Fibonacci sequences. The aperiodic deterministic sequences that have been almost exclusively considered so far deal with homogeneous layers. Introducing ZESST profiles therein allows the analysis of the interplay between smoothness and the scattering properties of the quasicrystals like pseudo band gaps and localized photonic states. Further work will be devoted to the nature of the "defects" that are deterministically inserted into the periodic sequence and the type of aperiodic sequence itself. Obviously, the ZESST profiles could also be used to study smooth periodic structures, i.e., photonic crystals.

With the ZESST profiles and more generally with the $\text{sech}(\hat{\xi})$-type profiles we now have at hand a *high-level* modeling tool that can be considered as an analytical Meccano able to fit to any (arbitrarily complex) graded index-profile and to easily provide the exact EM scattering properties.

Obviously, this new approach would save much effort as compared to the classical transfer matrix method, as already quoted earlier. Moreover, let us recall that analytical modeling of apodized Bragg filters or chirped Bragg reflectors, even after introducing the low-amplitude approximation, requires the implementation of a special function, namely the hypergeometric function [39, 60]. On the other side, another well-known technique, the coupled-mode method, is less demanding; yet, it is a perturbation-type theory that introduces various approximations; in particular, it is limited to low index modulation [30, 61]. A third method requires computing an infinitely nested set of integrals involving the logarithmic derivative of the admittance profile; again, for a practical implementation, only the first orders can be taken into account [31]. In contrast, with the ZESST profiles there is no restriction either on the index modulation height or on its rate: the EM-field computation is unconditionally exact. Assembling these S-form profiles and thereby multiplying the corresponding (high-level) transfer matrices, provides the scattering properties of the synthesized smooth multilayer without the discontinuity-induced artifacts that come along with the classical transfer matrix method with *constant* unit-cell profile.

This represents a new paradigm for modeling 1D graded index media and opens interesting perspectives for the inversion process, i.e., the design of refractive index profiles aimed at providing specified optical scattering properties.

Basically, the combination of the Liouville transformation and the PROFIDT method is applicable to any phenomenon that can be described by the following system of coupled first order ordinary differential equations:

$$\begin{cases} \dfrac{dF}{dz} = -i\omega g G \\ \dfrac{dG}{dz} = -i\omega f F, \end{cases} \quad (11)$$

where $f(z)$ and $g(z)$ are two real-valued, of same sign, $C^1$ functions representing the variable parameters (see [26]). By eliminating $G$ or $F$, a Helmholtz equation with variable coefficients is obtained for $F$, resp. $G$ (i.e. the so-called $\langle F \rangle$-form, resp. $\langle G \rangle$-form equation). The previous equations have been used to describe many types of evolutionary fields in physics. In the present paper, $F$ is for the electric field, $G$ is for the magnetic field, $f(z)$ is for $m_p \varepsilon(z) \cos^{1+m_p} \phi(z)$ and $g(z)$ is for $m_p \mu(z) \cos^{1-m_p} \phi(z)$. Other potential applications of the analytical tools are [26]: 1) electrical transmission lines with distributed inductance and conductance, 2) acoustic waves in a medium with graded mass density and sound velocity 3) longitudinal and shear elastic waves in a medium with graded mass density and elastic modulus, 4) ocean gravity waves. As such, the ZESST profile elements could be used for the analytical modeling of 1D *phononic* crystals and quasicrystals as well, which means structures with smooth periodic, resp. deterministic aperiodic, variations of the acoustic/elastic properties.

The results presented so far assumed that permittivity and permeability are real-valued and positive. Actually, to apply the standard Liouville transformation, they just have to be of the same sign. Hence, the PROFIDT method could also be used to model double-negative metamaterials (negative permittivity and permeability). Future work will be devoted to the exact analytical modeling of materials with complex-valued permittivity, i.e., complex-valued refractive index. One application is for materials with losses, another for materials with balanced gain and loss (PT-symmetric systems, i.e. systems unaffected after space-time reflection).

## APPENDIX A

The pseudoindex $n^*$ and the refractive index $n$, are related via the variable incidence angle $\phi(\xi)$ according to $n^* = n \cos^{m_p} \phi$, where $m_p = +1$ for the TE mode and $m_p = -1$ for the TM mode. Introducing the Snell-Descartes invariant $I_a = n_a \sin\phi_a = n(\xi)\sin\phi(\xi)$ and substituting $\phi(\xi)$ into the former relation, we get the two-way relations (see e.g. [1]):

$$TE: \begin{cases} n^* = \left(n^2 - I_a^{\,2}\right)^{1/2} \\ n = \left(n^{*2} + I_a^{\,2}\right)^{1/2} \end{cases} \quad (A1)$$

$$TM: \begin{cases} n^* = n^2 \left[n^2 - I_a^{\,2}\right]^{-1/2} \\ n = 2^{-1/2} n^* \left[1 \pm n^{*-1}\left(n^{*2} - 4 I_a^{\,2}\right)^{1/2}\right]^{1/2} \end{cases} \quad (A2)$$

In the latter equation, the plus or minus sign should be applied when the angle $\phi(\xi)$ is lower, resp. higher than 45°.

## APPENDIX B

The four entries of the analytical transfer matrix $\mathbf{M}_{\langle E \rangle}$ for a profile element solvable in the Liouville space are given by [26]:

$$\begin{bmatrix} A & B \\ C & D \end{bmatrix} =$$

$$\frac{1}{\Delta} \begin{bmatrix} s_0^{-1} s_1 (G - \mu_1 I) & im_p k_0 s_0^{-1} s_1^{-1} I \\ im_p k_0^{-1} s_0 s_1 (-\mu_0 G - \mu_1 H + \mu_0 \mu_1 I + J) & s_0 s_1^{-1} (-H + \mu_0 I) \end{bmatrix} \quad \textbf{(B1)}$$

where the subscripts 0 and 1 indicate that the corresponding functions are evaluated at $\xi = 0$, resp. at $\xi = \xi_1$; $\mu_{0,1} = s'_{0,1}/s_{0,1}$ and $G$, $H$, $I$, $J$, $\Delta$ involve the values taken at the two-layer edges by the independent functions $K(\xi, k_0)$ and $P(\xi, k_0)$ used to express the E and H fields and by their derivatives:

$$G = K_0 P_1' - K_1' P_0 \qquad H = K_0' P_1 - K_1 P_0'$$
$$I = K_0 P_1 - K_1 P_0 \qquad J = K_0' P_1' - K_1' P_0' \quad \textbf{(B2)}$$
$$\Delta = K_1 P_1' - K_1' P_1 = K_0 P_0' - K_0' P_0.$$

It is easy to verify that $\mathbf{M}_{\langle E \rangle}$ matrix is unimodular (i.e., its determinant is 1). Based on the expressions of the functions $K(\xi, k_0)$ and $P(\xi, k_0)$ related to the $\mathrm{sech}(\hat{\xi})$-type profiles (see Eq. (10)), the four terms $G$, $H$, $I$, $J$ become:

$$\begin{bmatrix} G \\ H \\ I \\ J \end{bmatrix} = -2i \begin{bmatrix} k_0^2 - \sigma_1^2 + \sigma_0 \sigma_1 & \sigma_0 (k_0^2 - \sigma_1^2) - \alpha^2 \sigma_1 \\ -k_0^2 + \sigma_0^2 - \sigma_0 \sigma_1 & \sigma_1 (k_0^2 - \sigma_0^2) - \alpha^2 \sigma_0 \\ \sigma_1 - \sigma_0 & \alpha^2 + \sigma_0 \sigma_1 \\ (\sigma_1 - \sigma_0)(k_0^2 + \sigma_0 \sigma_1) & \alpha^2 \sigma_0 \sigma_1 + (\sigma_0^2 - k_0^2)(\sigma_1^2 - k_0^2) \end{bmatrix} \quad \textbf{(B3)}$$
$$\times \begin{bmatrix} \alpha \cos(\alpha \xi_1) \\ \sin(\alpha \xi_1) \end{bmatrix}$$

where $\alpha \equiv \sqrt{k_0^2 - \xi_c^{-2}}$, whereas $\sigma_0$, $\sigma_1$ are the values taken at the layer edges by $\xi_c^{-1} \tanh(\hat{\xi})$. In addition, $\Delta = -2i\alpha k_0^2$.

We see that the four entries of the transfer matrix are fully determined by the following series of parameters: the end-values of the $s(\xi)$ profile, i.e., $s_0$ and $s_1$, the end slopes $s'_0$ and $s'_1$, the reduced wavenumber $k_0 \xi_1$, or equivalently $\xi_1 / \lambda$. The two other parameters $\xi_1/\xi_c$ and $\gamma$ that intervene in Eq. (B3) via $\sigma_0$, $\sigma_1$ and $\alpha$ are directly determined from the knowledge of $s_0$, $s_1$, $s'_0$ and $s'_1$. In the case of ZESST profiles, the derivatives $s'_0$ and $s'_1$ are zero, thereby implying that five terms in the expressions of the four entries in Eq. (B1) vanish.

The complex amplitude reflectance $r$ and the intensity reflectance $R$ of a layer (or a multilayer), as represented in Fig. 1, are given by:

$$R = |r|^2, \quad r = \frac{m_p n_a^* - Y}{m_p n_a^* + Y}, \quad Y = \frac{C + D m_p n_s^*}{A + B m_p n_s^*}, \quad \textbf{(B4)}$$

where $A$, $B$, $C$, $D$ are the entries of the transfer matrix of the layer (or multilayer), and $n_a^*$ and $n_s^*$ are the pseudo-indices of the incident medium, resp. substrate [1].

**Acknowledgment**. This work was funded within the framework of the RGC-DSG Program: "*Codes optiques*". The author is grateful to Grégory Vincent, Yves-Michel Frédéric and Grégoire Ky for valuable discussions.